\documentclass[11pt]{article} 
\usepackage{graphicx} 
\usepackage{parskip}
\usepackage[margin=1in]{geometry}
\usepackage{amsmath}
\usepackage{caption}
\usepackage{setspace}
\usepackage[left]{lineno}

\DeclareCaptionLabelSeparator{pipe}{\;\textbar\;}
\title{Epidemic amplification by correlated superspreading}

\author{%
  \mbox{Noah Silva de Leonardi}\thanks{%
    \mbox{Institute for Health Metrics and Evaluation, Hans Rosling Center for Population Health, University of Washington.}%
  }%
  \and
  \mbox{Benjamin D. Dalziel}\thanks{%
    \mbox{Departments of Integrative Biology and Mathematics, Oregon State University.}\newline
    Email: \texttt{benjamin.dalziel@oregonstate.edu}%
  }%
}

\begin{document}
\maketitle 
\onehalfspacing\begin{abstract}
Infectious pathogens often propagate by superspreading, which focusses onward transmission on disproportionately few infected individuals\cite{woolhouse1997heterogeneities,lloyd2005superspreading,althouse2020superspreading,lau2020characterizing}.
At the same time, infector-infectee pairs tend to have more similar transmission potentials than expected by chance, as risk factors assort among individuals who frequently interact\cite{newman2002assortative, salje2016social, cevik2021networks}.
A key problem for infectious disease epidemiology, and in the dynamics of complex systems, is to understand how structured variation in individual transmission will scale to impact epidemic dynamics\cite{ferrari2006network, bansal2007individual, vespignani2009predicting, balcan2011phase, metcalf2015seven, pellis2015eight}.
Here we introduce a framework that reveals how population structure shapes epidemic thresholds, through autocorrelation of individual reproductive numbers along chains of transmission.
We show that chains of superspreading can sustain epidemics even when the average transmission rate in the host population is below one, and derive a mathematical threshold beyond which correlated superspreading allows epidemics in otherwise subcritical systems.
Empirical analysis of $47$ transmission trees for $13$ human pathogens indicate self-organizing bursts of superspreading are common and that many trees are near the critical boundary.
Vaccination campaigns that proceed up assortative hierarchies of transmission are predicted to sustain the force of infection until herd immunity is reached, providing a mechanistic basis for threshold dynamics observed in real-world settings\cite{herzog2011heterogeneity,barclay2014positive,glasser2016effect,robert2022impact}.
Conversely, modulating correlations in transmission, rather than mean or variance, could enable cities and other complex systems to develop immune-like capacities that suppress contagion while preserving core functions.
\end{abstract}

\clearpage
\section*{}
Contagion processes play a central role in population biology and in the dynamics of complex systems.
This is evident in epidemics of infectious disease\cite{kermack1927contribution, diekmann1990definition, anderson1992infectious}, but also in the spread of advantageous alleles\cite{fisher1937wave,lieberman2005evolutionary, markov2023evolution}, and in the propagation of influence in social systems\cite{kempe2003maximizing, saberi2020simple, kawakatsu2021emergence}.
Outbreaks spread unevenly in complex populations\cite{han2023transmission}, as individual transmission rates vary due to physiological differences influencing pathogen load and shedding\cite{devincenzo2010viral, cheng2004viral}, social contact patterns that shape who acquires infection from whom\cite{adam2020clustering, cauchemez2011role}, and physical conditions that modulate pathogen persistence and dispersal range in the environment\cite{shaman2010absolute, baker2024increasing}.
Transmission heterogeneity in turn causes superspreading, where a majority of new infections are acquired from individuals whose transmission rate is far above the population average\cite{woolhouse1997heterogeneities,lloyd2005superspreading}. 

The impact of transmission heterogeneity depends on how it is organized.
If individual transmission rates are statistically independent, then superspreading does not affect the risk or trajectory of large epidemics, as individual variation in transmission potential averages out at the population level\cite{lloyd2005superspreading}. 
By contrast, non-independent transmission rates can change the conditions under which a local outbreak evolves into a major epidemic\cite{newman2002assortative, feld1991your, pastor2001epidemic,volz2011effects,colizza2007invasion,leventhal2015evolution}.
This is illustrated by compartmental models that subdivide a host population into classes (e.g., by age, or spatial location), where the epidemic threshold corresponds to the spectral radius of the next-generation matrix, which is constructed from the inter-class transmission rates\cite{diekmann1990definition,colizza2008epidemic,diekmann2010construction,manna2024generalized}.
In network epidemic models, where individual transmission potentials are explicitly wired together, the epidemic threshold is lower in networks with higher degree variance and assortativity\cite{newman2002spread, newman2002assortative, pastor2001epidemic}.
Decades of work link host population structure to epidemic dynamics\cite{diekmann1990definition, grenfell2001travelling,newman2002spread,colizza2007invasion,dalziel2014contact,dalziel2018urbanization}, and epidemic thresholds can be calculated for any host population and pathogen, subject to data availability and model constraints, using generalized methods such as next-generation matrices\cite{diekmann1990definition} or generating-function approaches\cite{newman2002spread}. 
Crucially, however, these methods do not provide a predictive understanding of how individual variability is amplified or attenuated by population structure.
An open question that unites modeling approaches is how organizing heterogeneity propagates to determine epidemic thresholds\cite{pastor2001epidemic,istvan2017mathematics,sun2021transmission,manna2024generalized}.
This question underlies important cross-scale challenges including predicting the impact of layered control measures\cite{sun2021transmission}, forecasting how demography and climate will combine to impact epidemic risk\cite{baker2022infectious}, and incorporating human behavior into epidemiological models\cite{funk2015nine}.

Here we quantify how organizing heterogeneity alters epidemic thresholds by modeling autocorrelation of individual reproductive numbers along chains of transmission, in both simulated and real outbreaks. 
At base, epidemics of infectious disease are possible if and only if at least one additional host is expected to acquire the infection from an individual who is currently infected\cite{kermack1927contribution, anderson1992infectious}.
Crucially, this expectation is a ``pathwise'' property, operating along chains of transmission\cite{brockmann2013hidden}, rather than a simple expectation for a randomly-selected individual\cite{bansal2007individual}.
This suggests that organized chains of superspreading could scale up outbreaks and sustain epidemics, while the mean transmission potential averaged over the population is prohibitively low\cite{kim2025superspreading}, as could be the case where landscapes of susceptibility significantly modulate spread, or in populations near critical vaccination coverage\cite{biggerstaff2014estimates,graham2019measles}.
However the role of organized superspreading in sustaining transmission remains poorly understood, and its contribution to real-world epidemic dynamics is largely unknown.

Let $z_i$ represent the number of onward transmission events that occur if an individual $i$ becomes infected.
Following standard theory\cite{lloyd2005superspreading}, we assume $z_i$ arises from a Poisson process conditioned on $i$'s individual reproductive number $\nu_i$ which represents their average transmission potential over many realizations of an outbreak in which they become infected. 
The individual reproductive number is itself a random variable, reflecting systematic variation among individuals in transmission potential, and is gamma distributed,
\begin{align}
z_i \mid \nu_i &\sim \mathrm{Poisson}\left(\nu_i\right) \\
\nu_i &\sim \mathrm{gamma}(\bar{\nu}_i, k)
\end{align}
with mean $\bar\nu_i$ and dispersion $k$. 
As a gamma mixture of Poissons, the overall distribution of $z_i$ (the ``ofspring distribution'' in the language of branching process models) is negative binomial, with mean $\bar\nu_i$ and variance $\bar\nu_i + \bar\nu_i^2 / k$. 
The expected number of secondary infections acquired from the first infected individual is given by the basic reproductive number in the population, $\bar\nu_0 = \bar\nu = R_0$. 
The standard model for superspreading\cite{lloyd2005superspreading} further assumes that individual reproductive numbers are independent and identically distributed, thus $\bar\nu_i = \bar\nu$ for all $i$.

We extend the standard model by allowing $\bar\nu_i$ to change along chains of transmission as
\begin{equation}
\bar\nu_i = \bar\nu + \delta \left(\nu_{p(i)}  - \bar\nu\right)
\end{equation}
where $p(i)$ returns the index of the infection prior to $i$ in a transmission chain. 
The autocorrelation parameter $\delta$ represents the propensity for infector-infectee pairs to have more similar reproductive numbers than would be expected if they were independently distributed, due to shared conditions that influence transmission rates.
Setting $\delta=0$ recovers the standard model as a special case, while $\delta>0$ introduces correlated superspreading, without changing the population-wide mean and dispersion in individual reproductive numbers, given by $\bar\nu$ and $k$ respectively.
While this approach is flexible with respect to the sources of transmission heterogeneity, it is precise in its mechanism of action---addressing population structure as experienced by the pathogen as it spreads.

Our results show that even weak path-level organization of transmission potential ($0 < \delta < 0.1$) can generate outbreaks that are orders of magnitude larger than if superspreading occurred at random (Figure 1).
For instance, when weak correlation ($\delta = 0.08$) is introduced to a population with subcritical mean transmission rate ($\bar\nu = 0$) and marginal levels of superspreading ($k = 1$), the final size of a 1-in-100 outbreak (99th percentile) increases by approximately 10 fold, from about $100$ to $1000$ cases.
These effects intensify as dispersion increases, within empirically observed ranges\cite{lloyd2005superspreading,taube2022open}.
As correlation in superspreading increases, the tail distribution of outbreak sizes places more mass in extreme events and decays more slowly, indicating potential for very large outbreaks.
The emergence of very large outbreaks in subcritical regimes shows that path organization substantially alters the risk posed by superspreading. 
In particular, superspreading in social-like systems is far more dangerous than the standard model indicates\cite{lloyd2005superspreading, althouse2020superspreading}.

\begin{figure}[ht]
  \centering
  \includegraphics[width = \textwidth]{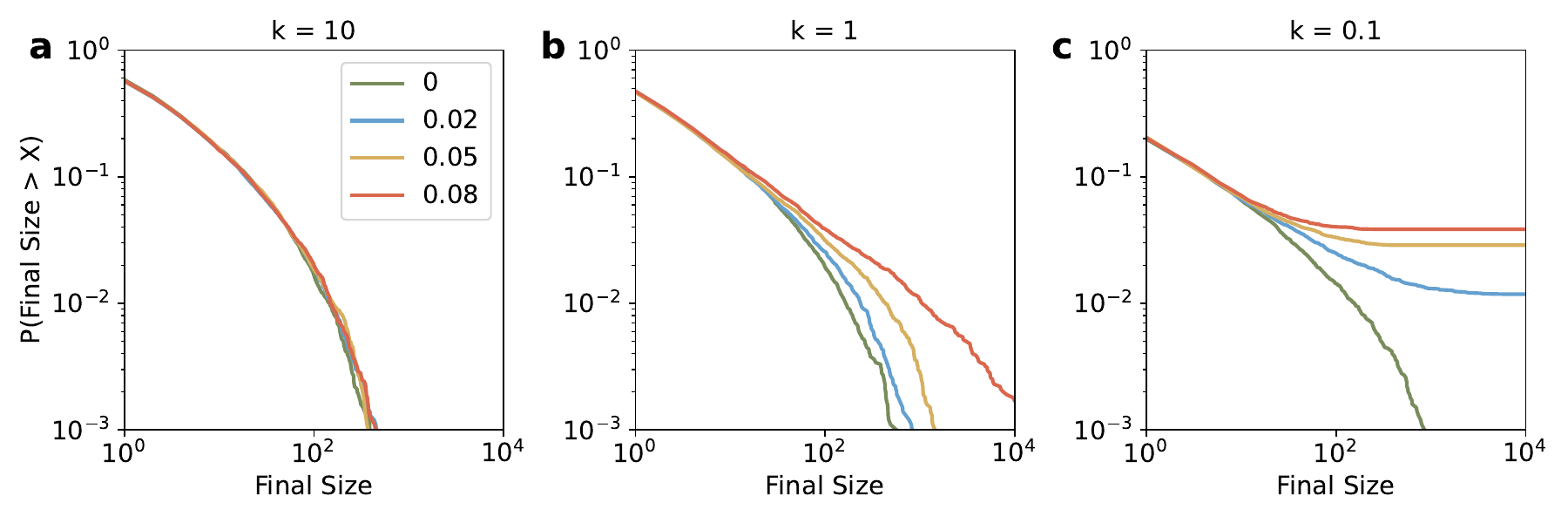}
  \caption{\small{\textbf{Correlated superspreading increases the final size and tail risk of outbreaks}. Each curve shows the tail distribution of $10^4$ simulated outbreaks, each initiated with a single infected individual with reproductive number drawn from $\mathrm{gamma}(\bar\nu=0.9, k)$. Colors correspond to different values of the autocorrelation parameter $\delta$. Panels \textbf{a-c} correspond to different values of the dispersion parameter $k$, with lower values corresponding to higher levels of dispersion. Simulations were right censored at $10^4$ cases.}}\label{final_size}
  \noindent\rule{\textwidth}{0.4pt}
\end{figure}

We identify a critical threshold beyond which correlated superspreading sustains epidemics in otherwise subcritical regimes (i.e., when $\bar\nu < 1$), given by
\begin{equation}
  \delta^* = \frac{1 - \bar\nu}{1 - \bar\nu + \frac{1}{k}}
\end{equation}
where $\delta^*$ is level of autocorrelation above which local outbreaks have $P[\text{fadeout}] < 1$.
The threshold is derived mathematically (Methods), and confirmed by simulation (Figure 2).
As $\bar\nu \to 1$ from below, even minimal correlation allows epidemics.
As $\bar\nu \to 0$, epidemics remain possible under extreme, highly organized variation.
When $\bar\nu > 1$ and $\delta>0$, high dispersion can produce faster-than-exponential growth with limited predictability and many fadeouts.
Autocorrelation thus shapes the impact of individual variation on outbreak size, epidemic risk, and the limits of population-wide predictability, addressing the longstanding debate on when individual heterogeneity matters\cite{bansal2007individual}.

\begin{figure}[ht]
  \centering
  \includegraphics[width = \textwidth]{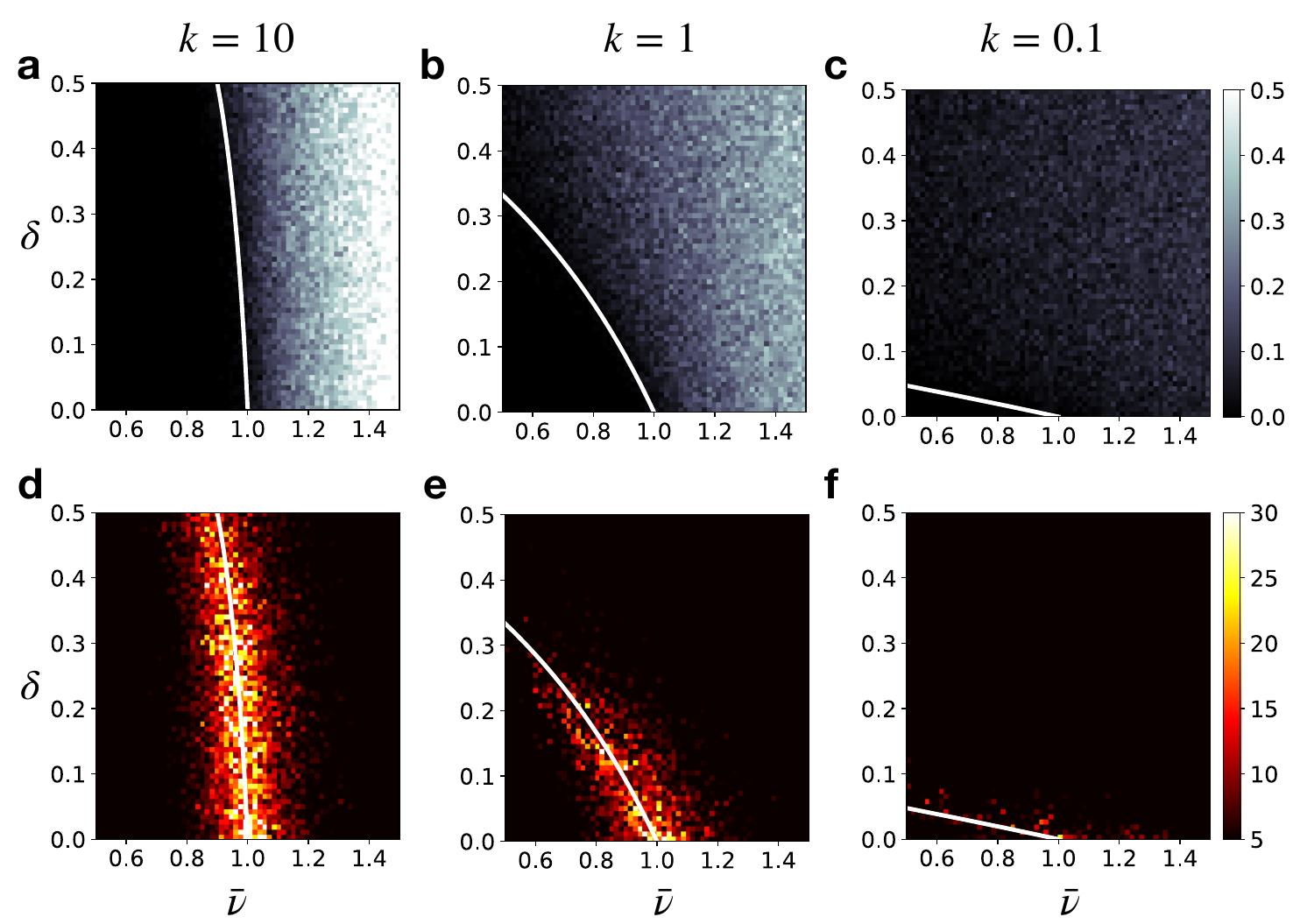}
  \caption{\small{\textbf{There is a critical threshold beyond which correlated superspreading can sustain epidemics in otherwise subcritical systems.} \textbf{a-c,} The proportion of simulated outbreaks reaching $10^3$ cases as a function of mean transmission rate ($\bar\nu$) and path-correlation in transmission potential ($\delta$).
  Dispersion lowers the critical level of organization required for superspreading to drive epidemics, but increases the risk that local outbreaks will fade out\cite{lloyd2005superspreading}. Epidemic risk in biological, social and technological systems may therefore be maximized at intermediate levels of dispersion, where there is sufficient variance to scale superspreading through organized transmission (\textbf{b}), but not so much variance that fadeouts dominate (\textbf{c}). \textbf{d-f,} The average size of outbreaks that did not reach $10^3$ cases, with larger outbreaks marking the critical boundary. The white line shows the analytical threshold $\delta^* = (1-\bar\nu) / (1-\bar\nu + k^{-1})$. Each pixel shows the result of 100 replicate simulations.}}\label{surfaces}
  \noindent\rule{\textwidth}{0.4pt}
\end{figure}

Turning to empirical data, we confront the null-hypothesis that $\delta$ is effectively $0$ in real-world systems.
This is plausible: while successive superspreading events have been observed\cite{taube2022open,jang2020cluster}, transmission chains often traverse biological, social, and environmental contexts\cite{faye2015chains}, which may erode similarities in transmission rates among infector-infectee pairsx.
Given successive values of $z$ observed along transmission chains, our model specifies a likelihood function that allows parameter estimation by Markov chain Monte Carlo, yielding posterior distributions for $\delta$, and other model parameters (Methods).
We fitted our model to a database of empirical transmission trees for human pathogens assembled by Taube et al.\cite{taube2022open}, using all trees that had $\geq 3$ generations and finite estimates of the dispersion parameter $k$, which yielded $47$ trees spanning $13$ human pathogens (Methods). 
This sample should not be considered representative, but can reject our null-hypothesis by counterexample.

Over two thirds of trees had posterior mean values for $\delta$ that exceeded $0.1$, which is sufficient to substantially alter epidemic thresholds under realistic values for $k$ and $\bar nu$ (Fig. 1, 2).
The median of per-tree posterior means was 0.126, (range $0.0040$ to $0.3373$) and the median of the first quartiles of the posterior distributions for each tree was $0.0393$ (range $0.00171$ to $0.08640$).
Thus, across the empirical trees in the database, posterior distributions concentrate substantially away from zero, and the null hypothesis that $\delta=0$ is not supported.

Estimates for $\delta$ straddle the critical region where subcritical systems can be tipped into epidemic spread by correlated superspreading (Fig. 3d).
Interestingly, transmission trees for SARS-CoV-2 appeared to occupy the central region of $\delta-k$ parameter space where superspreading can scale outbreaks without fadeouts dominating, as described above (Fig. 2, 3d,e).
Since estimates of $\delta$ are substantial in empirical trees, can be readily estimated from contact tracing data, and determine the risk associated with a given level of dispersion $k$, it follows that $\delta$ should, where possible, be estimated and reported alongside other key epidemiological parameters in outbreak investigations and studies of emerging infectious disease threats.

\begin{figure}[ht]
  \centering
  \includegraphics[width = \textwidth]{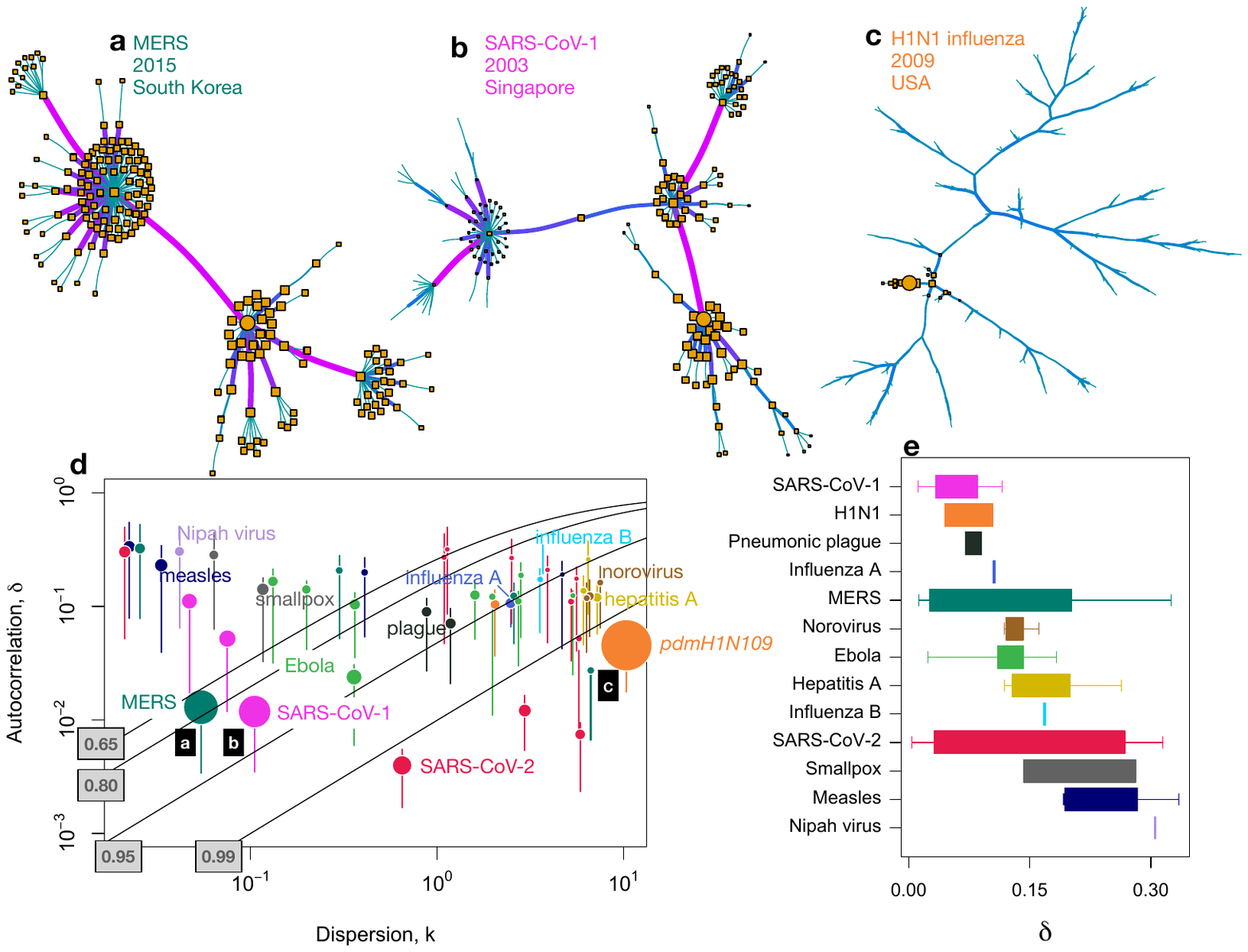}
  \caption{\small{\textbf{Empirical transmission trees show a range of autocorrelation in individual transmission rates.} 
  \textbf{a-c,} Empirical trees assembled by~\cite{taube2022open}: Middle East Respiratory Syndrome Coronavirus (MERS) in Korea in 2015\cite{zhang2017estimating}, Severe acute respiratory syndrome (SARS) in Singapore in 2003\cite{leo2003severe}, and H1N1 influinza in the United States in 2009\cite{cauchemez2011role}. 
  Circles show the first case. 
  Successively smaller squares label the first 5 generations. 
  The thickness and color of the edges is proportional to the product of the number of cases associated with their vertex pair, so edges connected successive superspreading events are wider. 
  \textbf{d} Postior mean estimates of $\delta$ and $k$ for each tree. 
  Vertical lines enclose the interquartile range of the posterior distribution. 
  Curves show the critical thresholds for the respective values of $\bar\nu$ in the grey boxes.
  Marker size is proportion to the number of cases in the tree range from $8$ to 286 (median, mean, s.d.: 32, 43.3617, 50.8798). The points are color-coded by pathogen. \textbf{e}, distrubutions across pathogens. 
  }}\label{empirical}
  \noindent\rule{\textwidth}{0.4pt}
\end{figure}

We examined the implications of correlated superspreading for the design of control systems.
This required linking our framework to susceptible dynamics.
To do so, we used a compartmental model structured by individual reproductive number, such that $n$ secondary infections are expected to be acquired from an individual in the $n$th class (Methods).
The next-generation matrix for this model has entries $\mathrm{K}_{nm} = mP[n|m]$, where $P[n|m]$ is the probability that a host of type $n$ acquires infection from a host of type $m$. 
The conditional probability $P[n|m]$ is parameterized as negative binomial distribution with mean $\bar\nu + \delta(m-\bar\nu)$ and dispersion $k$, consistent with our framework.
It is worth noting that a compartment model structured by individual reproductive number illustrates why autocorrelation determines the impact of individual variation on epidemic dynamics. 
When $\delta=0$ then the columns of $\mathrm{K}$ differ only by a scalar multiple $m$, and its spectral radius is the mean transmission potential in the population $\bar\nu$, regardless of individual variation ($k$).
Conversely $\delta>0$ implies that the spectral radius emerges from the full structure of the matrix, allowing $k$ to influence epidemic thresholds (Methods).

We used the compartment version of our model to consider the impact of correlated superspreading on vaccination campaigns.
In practice, vaccination rollouts often reach lower-transmission groups first, as individuals with high transmission potential tend to face barriers to access to healthcare, and vaccination programs may prioritize severe disease risk over transmission risk.
We therefore simulated vaccination as sequentially targeting classes in ascending order of $n$, each to critical coverage $1-1/R_0$, with $R_0$ given by the spectral radius of $K$.
Classical theory assumes vaccination is done at random and predicts the population reproductive number after control is applied $R_c$ will decline linearly with coverage, $R_c=(1-c)R_0$, where $c$ is the proportion vaccinated. 
Under these conditions correlated superspreading did not change the dynamics.
Sequential rollout instead produces a nonlinear response, where $R_c$ initially declines more slowly until higher-$n$ classes are reached, then drops more rapidly.
This nonlinear effect intensifies as $\delta$ increases, producing threshold-like dynamics where $R_c \approx R_0$ for much of the rollout before collapsing abruptly (Fig. 4).
These results provide a quantitative basis for predicting threshold behavior observed in real-world vaccination campaigns\cite{herzog2011heterogeneity,barclay2014positive,glasser2016effect,robert2022impact}.

We also explored a proof-of-concept for a new class of intervention in which $\delta$ is used as a control variable.
For example, this could correspond to an ``intelligent city'' that adaptively adjusts mobility patterns through incentives so that transmission chains are less likely to connect among high-transmission groups\cite{eubank2004modelling}. 
To test this idea quantitatively, we coupled our compartment model to a proportional–integral–derivative (PID) controller that dynamically reduced $\delta$ in response to incidence. 
Short-term reductions in $\delta$ in response to rising incidence substantially blunted epidemic peaks, achieving performance comparable to that of a seasonal influenza vaccination program in the United States during a typical year (Fig. 4b).
In practice, the efficacy of $\delta$-based control will likely depend on many factors, and these types of feedback loops can produce complex system behaviors which are not yet fully understood\cite{weitz2020awareness}.

\begin{figure}[ht]
  \centering
  \includegraphics[width = \textwidth]{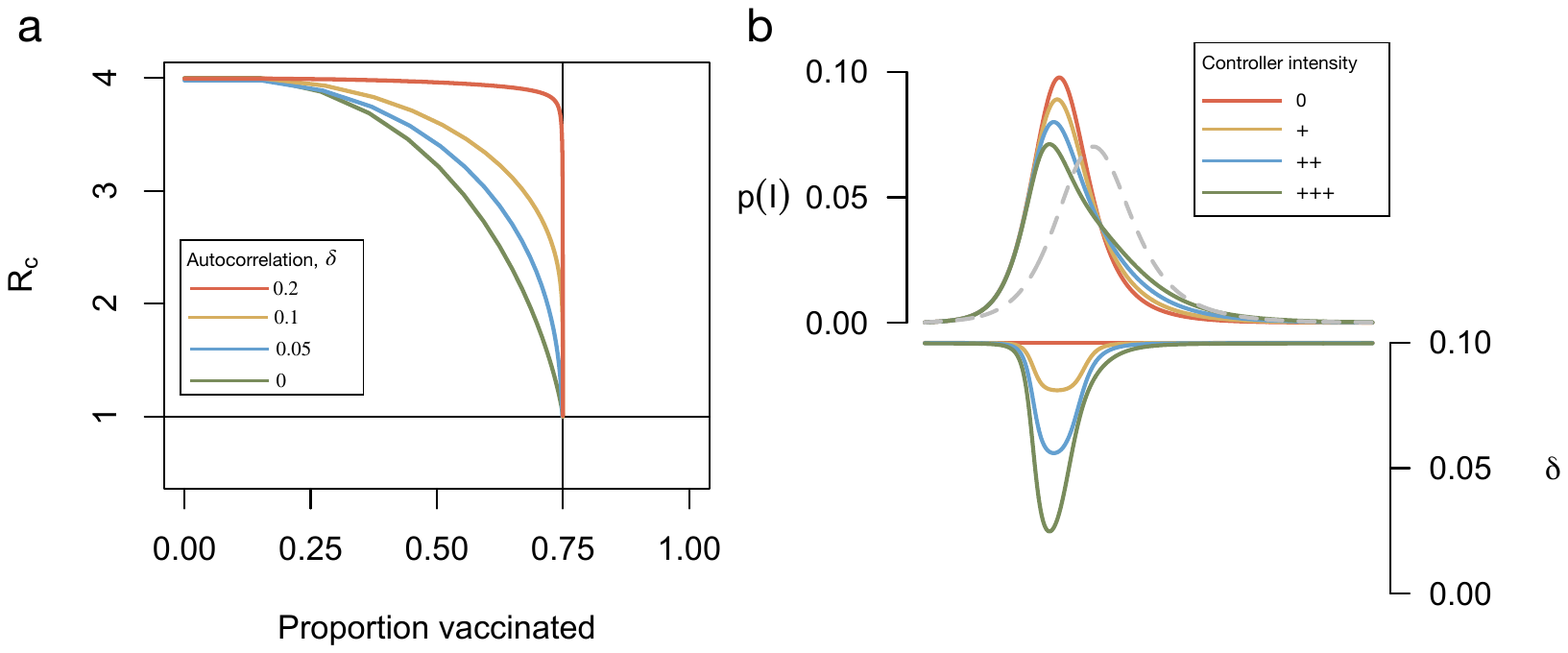}
  \caption{\small{\textbf{Implications of correlated superspreading for the design of control systems.}
\textbf{a}, The response of the controlled reproductive number, $R_{\mathrm{c}}$, to vaccination depends on the correlation among individual reproductive numbers along transmission chains, $\delta$. When vaccination rollouts ascend transmission hiearchies, increasing $\delta$ creates threshold dynamics, with $R_{\mathrm{c}} \approx R_0$ until herd immunity is reached. Dispersion is fixed at $k=1$, and mean transmission is scaled so that $R_{0}=4$ across $\delta$ values. The population is structured by individual reproductive number and classes are vaccinated to the critical threshold ($1-1/R_0$) in ascending order. 
\textbf{b}, Epidemic dynamics under $\delta$-based feedback control. Top: proportion of the population infected over time; bottom: $\delta(t)$ adjusted by a PID controller driven by incidence. Stronger control intensities can efficiently reduce epidemic peaks, by providing a rapid, self-limiting response. As a counterfactual the dashed line simulates a vaccination program comparable to seasonal influenza vaccination in the United States with coverage $v=0.45$~\cite{cdc2024flu} and efficacy $e=0.20$~\cite{grijalva2024estimated}, giving $R_c = (1-ve)R_0$. Controller parameters are $K_P=10^{-4}$, $K_I=0$, $K_D=0$. Other parameters approximate seasonal influenza: $\bar \nu=1.2$, $k=1$.}}\label{control}
  \noindent\rule{\textwidth}{0.4pt}
\end{figure}

Highly variable transmission rates have been observed during emergence, prolonged tails, and surge phases of epidemics\cite{riley2003transmission, blumberg2013inference, lau2017spatial, graham2019measles, aber2025time}.
Our results challenge the conventional view that this stochastic variation in individual transmission does not influence epidemic trajectories\cite{kermack1927contribution,anderson1992infectious,lloyd2005superspreading}, by showing that correlated bursts of transmission, which are present in empirical transmission trees (Fig. 3), can ``bend'' epidemic thresholds (Fig. 2).
This extends the principle that stochastic fluctuations may influence the dynamical structure of complex adaptive systems\cite{prigogine1978time,grenfell1998noise}.
Parallel work in evolutionary theory shows the evolution of pathogen transmissibility during subcritical outbreaks can accelerate emergence\cite{antia2003role}. 
Similarly, experiments with clonal populations of \emph{Escherichia coli} demonstrate that random variation in reproductive rates among lineages accelerates population growth\cite{hashimoto2016noise}.

Epidemics will be most likely to emerge, and hardest to control, when transmission is both highly variable and highly organized.
These conditions are becoming more common as climate change interacts with demographic shifts such as urbanization\cite{dalziel2018urbanization,baker2022infectious}.
Adaptively reducing correlations in transmission---for example, by incentivizing small changes that desynchronize peak contact—--offers a new strategy for limiting transmission that is distinct from targeting mean or variance in transmission rates, and suggests how social-technological networks could modulate contagion at scale while preserving core functions.

\clearpage
\section*{Methods}
\subsection*{Critical threshold}
The model in equation 3 specifies a stochastic branching process where reproduction rates may be correlated along lineages.
This differs from the classical Galton–Watson process where reproductive rates are independent and identically distributed.
The threshold for persistence in a Galton–Watson process is  
\begin{equation}
\Pr[\text{extinction}] < 1 \;\;\Longleftrightarrow\;\; \mathrm{E}[z] > 1 ,
\end{equation}
where $\mathrm{E}[z]$ is the expected number of new individuals that arise from a randomly selected individual. 
The event ``extinction’’ occurs if the process eventually terminates, so that the total number of individuals ever produced is finite.
The complementary event, ``persistence’’ means that the population grows without bound. 
In the context of infectious disease epidemics, $z$ represents the number of secondary infections acquired from a randomly selected infected individual, and persistence corresponds to the occurrence of a large epidemic.  

As described in the main text we define the expected number of infections acquired from a particular infected individual $i$ by their reproductive number $\nu_i$
\[
\mathrm{E}[z_i] \;=\; \nu_i.
\] 
Equation 3 provides the expectation for $\nu_i$ conditioned on the individual reproductive number of the previous infection in the transmission chain, $\nu_{p(i)}$
\[
E[\nu_i \mid \nu_{p(i)}] = \bar \nu + \delta(\nu_{p(i)} - \bar\nu) = \bar \nu_i
\]
By the law of total expectation,  
\begin{align} 
\mathrm{E}[z] 
&= \mathrm{E}\!\left[\mathrm{E}[z_i]\right] \notag\\
&= \mathrm{E}[\nu_i] \notag\\
&= \mathrm{E}\!\left[\,\mathrm{E}[\nu_i \mid \nu_{p(i)}]\,\right] \notag\\
&= \mathrm{E}[\bar\nu_i] \notag\\
&= (1-\delta)\,\bar\nu \;+\; \delta\,\mathrm{E}[\nu_{p(i)}]
\end{align}  
where $\bar\nu$ is the mean transmission potential averaged over the entire host population, which governs the asymptotic distribution of chain averages, though individual chains may fluctuate around it.
Indeed, where $\bar\nu < 1$ but $\delta > 0$, local bursts of transmission may sustain global persistence, with the asymptotic growth rate of the epidemic driven by stochastic fluctuations, a phenomenon evocative of fluctuation-driven organization in systems far from thermodynamic equilibrium, with cities, and other living systems, as classic examples\cite{prigogine1978time}.

Setting the lefthand side of eqn (6) equal to 1 reveals that the critical boundary occurs along a curve in the $\delta-\bar\nu$ plane:
\begin{equation}
\delta^* = \dfrac{1 - \bar \nu}{\mathrm E{\left[\nu_{p(i)}\right]}^* - \bar \nu}
\end{equation}
where $E{\left[\nu_{p(i)}\right]}^*$ denotes the expected value at the critical boundary.

Deriving an expression for ${\mathrm{E}[\nu_{p(i)}]}^*$ is the next step. 
Let $\nu_j$ denote the reproductive number of an individual $j$ who was infected in the generation prior to $i$, so that it is possible that $i$ acquired the infection from $j$.
Importantly, the distribution of $\nu_j$ is not generally the same as the distribution of $\nu_{p(i)}$, because the probability that $i$ acquired the infection from $j$ is proportional to the total number of secondary cases acquired from $j$, $z_j$, which biases $\nu_{p(i)}$ toward higher-transmission individuals.
Recalling that case counts satisfy $z_j \sim \mathrm{Poisson}(\nu_j)$ and are independent conditional on $\nu_j$, and using the fact that independent Poisson counts are distributed multinomially proportional to their rates, the probability that $i$ acquired the infection from $j$ is
\[
\mathrm{P}\!\left[p(i)=j \,\middle|\, \nu_1,\dots,\nu_m\right]
= \frac{\nu_j}{\sum_{\ell=1}^m \nu_\ell}.
\]
where $m$ is the number of individuals in the generation previous to $i$.

Then,
\begin{align}
\mathrm{E}\!\left[\nu_{p(i)} \,\middle|\, \nu_1,\dots,\nu_m\right]
&= \sum_{j=1}^m \nu_j \,\mathrm{P}\!\left[p(i)=j \,\middle|\, \nu_1,\dots,\nu_m\right] \notag \\
&= \sum_{j=1}^m \nu_j \,\frac{\nu_j}{\sum_{\ell=1}^m \nu_\ell} \notag \\
&= \frac{\sum_{j=1}^m \nu_j^2}{\sum_{j=1}^m \nu_j} \notag \\
&= \frac{\tfrac{1}{m}\sum_{j=1}^m \nu_j^2}{\tfrac{1}{m}\sum_{j=1}^m \nu_j} \notag \\
&= \frac{\mu^2 + \sigma^2}{\mu} \notag
\end{align}
where $\mu = \tfrac{1}{m}\sum_{j=1}^m \nu_j$ is the empirical mean and $\sigma^2 = \tfrac{1}{m}\sum_{j=1}^m {\left(\nu_j-\mu\right)}^2$ is the empirical variance of individual reproductive numbers in the generation prior to $i$.

In the large-population limit, the empirical mean $\mu$ converges in probability to $\mathrm{E}[\nu_j]$ and the empirical variance $\sigma^2$ converges in probability to $\mathrm{Var}[\nu_j]$.
Further, because $\nu_j$ are gamma distributed with common dispersion $k$, $\mathrm{Var}[\nu_j] = \mathrm{E}{[\nu_j]}^2 / k$.

Using the law of total expectation and the continuous mapping theorem,
\begin{align}
\mathrm{E}\!\left[\nu_{p(i)}\right]  
= &~\mathrm{E}\!\left[\mathrm{E}\!\left[\nu_{p(i)} \,\middle|\, \nu_1,\dots,\nu_m\right]\right] \notag \\
= &~\mathrm{E}\!\left[\frac{\mu^2 + \sigma^2}{\mu} \right] \notag\\
= &~\mathrm{E}[\mu] + \mathrm{E}\!\left[\frac{\sigma^2}{\mu}\right] \notag\\
\xrightarrow[m~\to~\infty]{} &~\mathrm{E}[\nu_j] + \frac{\mathrm{E}[\nu_j]}{k} \notag \\
= &~\mathrm{E}[\nu_j]\left(1 + \frac{1}{k}\right) \notag
\end{align}
Finally, we assume that at the critical boundary $\mathrm{E}[\nu_j]=1$. Thus
\[
\mathrm{E}\!{\left[\nu_{p(i)}\right]}^* = 1 + \frac{1}{k}
\]

Substituting into equation (7) yields the critical threshold
\[
\delta^* = \dfrac{1 - \bar \nu}{1 - \bar\nu + \frac{1}{k}}
\]
which is corroborated by direct simulation of the branching process (Fig. 2).

\subsection*{Empirical inference}
For each tree we analyzed from~\cite{taube2022open} the data consist of the observed offspring count $z_i$ and the offspring count $z_{p(i)}$ of the previous infection in the transmission chain, for each non-root infection in the tree.
We fit our model to each empirical transmission tree using Hamiltonian Monte Carlo implemented in Stan\cite{carpenter2017stan}, via the rstan package in R\cite{standev2018rstan}.

The distribution of $z_{p(i)}$ conditioned on its individual reproductive number is
\[
z_{p(i)} \mid \nu_{p(i)} \sim \mathrm{Poisson}(\nu_{p(i)})
\]
with $\nu_{p(i)}$ treated as a latent parameter to be estimated. 

The distribution of $z_i$, conditional on the full set of parameters $\nu_{p(i)}$, $\delta$, $\bar\nu$, $k$, is negative binomial with mean $\mu_i \;=\; (1-\delta)\,\bar\nu \;+\; \delta\,\nu_{p(i)}$
and dispersion $k$,
\[
z_i \mid \nu_{p(i)}, \delta, \bar\nu, k \;\sim\; \mathrm{NegBin}(\mu_i,k).
\]
The joint likelihood is therefore the product of the Poisson factors for the parent counts and the negative binomial factors for the observed child counts. 
In Stan this is implemented using the \texttt{neg\_binomial}$(k, lambda)$ parameterization, where $\lambda = k/\mu_i$ ensures the expected value is $\mu_i$:
\[
\mathcal{L}(\delta,\bar\nu,k,\{\nu_{p(i)}\}\mid z, z_p)
= \prod_{i=1}^n f\!\left(z_{p(i)} \mid \nu_{p(i)}\right)
  \;\times\;
  \prod_{i=1}^n g\!\left(z_i \mid \nu_{p(i)}, \delta, \bar\nu, k\right).
\]
where $f(z_{p(i)} \mid \nu_{p(i)})$ is the probability mass function (PMF) of Poisson random variable with mean $\nu_{p(i)}$, and 
$g(z_i \mid \nu_{p(i)}, \delta, \bar\nu, k)$ is a negative binomial PMF with mean $\mu_i$ and dispersion $k$.

We placed weakly informative priors on all model parameters. The autocorrelation parameter was given a uniform prior,
\[
\delta \sim \mathrm{Beta}(1,1), \qquad 0 \leq \delta \leq 1.
\]
The population mean reproductive number had a gamma prior,
\[
\bar\nu \sim \mathrm{Gamma}(10,10),
\]
corresponding to a prior mean of 1 and variance of 0.1. The dispersion parameter was assigned a half-normal prior,
\[
k \sim \mathrm{Normal}^+(0,10),
\]
implemented as a normal distribution with mean 0 and standard deviation 10, truncated below at 0. Each latent parent reproductive number was also assigned a half-normal prior,
\[
\nu_{p(i)} \sim \mathrm{Normal}^+(0,100).
\]

We fit each tree separately using 4 chains × 10,000 iterations per chain with a 50\% warmup. 
To verify convergence we confirmed that there were no divergences or max treedepth saturations, 
that rank-normalized $\hat R \leq 1.01$ for all estimated parameters, and that visual inspection of traceplots showed rapid mixing across chains.

\subsection*{Vaccination}
To analyze the dynamics of susceptibility with correlated superspreading we use a compartmental epidemic model for a host population structured by individual reproductive number, such that individual $i$ in the $n$th class has $\nu_i = n$. 
Let $S_n$ and $I_n$ denote the number of susceptible and infectious individuals in each class, and $N_n$ the total population size of class $n$. 
We consider a pathogen that confers immunity following infection for a duration longer than the time scale of an epidemic. 
The number of individuals removed from the susceptible and infectious pools by prior infection is $N_n - S_n - I_n$, and the total population size is $N = \sum_n N_n$. 
Epidemic dynamics are given by the standard compartmental epidemic model for structured populations
\begin{align}
  \frac{dS_n}{dt} &= -S_n \sum_m \lambda_{nm} \notag\\
  \frac{dI_n}{dt} &= S_n \sum_m \lambda_{nm} - \gamma I_n \notag
\end{align}
where $\gamma$ is the rate of removal from the infectious class, and the force of infection is
\[
\lambda_{nm} = \beta_{nm} \frac{I_m}{N_m},
\]
which is the instantaneous hazard experienced by a susceptible in class $n$ from infectious individuals in class $m$. 
The who-acquires-infection-from-whom (WAIFW) matrix $\beta$ has entries $\beta_{nm} = m \mathrm{P}[n|m]$ 
with $\mathrm{P}[n|m]$ the probability that a new infection of class $n$ is acquired from a class $m$ individual. 
Consistent with our model, this is specified by the probability mass function of a negative binomial with mean $\bar\nu + \delta(m-\bar\nu)$ and dispersion $k$, normalized to account for truncation at a finite number of classes.
We use 128 classes to approximate the distribution and set the total population size to $N=10^6$.
We set $\gamma = 1$ without loss of generality, so time is measured in units equal to the mean generation time of the pathogen, so that the next-generation matrix is $\mathrm K = \beta~$\cite{diekmann2010construction}. 
Thus $R_0 = \rho(\beta)$, where $\rho$ denotes the spectral radius, which is in turn determined by $\delta$, $k$, and $\bar\nu$, via the distribution $\mathrm{P}$. 
We explore model dynamics by numerical integration, starting with 100 infectious individuals in class $2$. 
The initial distribution of susceptible individuals is set proportional to the dominant eigenvector of $\mathrm{K}$, approximating the stable class distribution near the disease-free equilibrium.

To model the impact of vaccination we assumed an idealized vaccine that provides complete protection against infection, and explicitly represent the structural impact of vaccination on transmission pathways through the next-generation matrix.
This facilitates spectral analysis of epidemic thresholds under vaccine-induced heterogeneity in populations structured by individual reproductive number.
In this framework, changes in the initial distribution of susceptibles due to vaccination arise through changes in the dominant eigenvector of $\mathrm{K}$.
Specifically, vaccination of class $n$ individuals reduces the probability that new infections are assigned to that class by a factor $(1-v_n)$, where $v_n$ is the proportion of individuals in class $n$ who are vaccinated.
The removed probability mass is reallocated to the $n=0$ class, consistent with individuals who as a result of vaccination do not become infectious following exposure.
Denoting this modified distribution by $P'[n \mid m]$, 
\[
P'[n \mid m] \;=\;
\begin{cases}
P[0 \mid m] + \displaystyle\sum_{r} v_r\,P[r \mid m], & n=0, \\[1.2em]
(1-v_n)\,P[n \mid m], & n>0,
\end{cases}
\]
The next-generation matrix under vaccination is then
\[
K'_{nm} = m \, P'[n \mid m].
\]

\subsection*{Adaptive control}
The PID controller drives a nonlinear update of $\delta$ in response to incidence.
Incidence is measured as the total number of infectious individuals $I = \sum_n I_n$. 
The controller state $x(t)$ evolves according to a proportional (P) update rule,
\[
\frac{dx}{dt} = -k_P \sum_n \frac{dI_n}{dt},
\]
where $k_P>0$ is the control parameter. Thus incidence velocity drives controller velocity, such that increases in incidence cause negative increments in $x(t)$.

The controller state is linked to epidemic dynamics through a bounded sigmoid function,
\[
u(x) \;=\; (1-a) \;+\; \frac{a}{1 + e^{-b(x+c)}},
\]
where $a > 0$ sets the range of the control response, $b > 0$ determines the slope, and $c$ shifts the activation threshold. 
Results in the Fig. 4 used a = (0, 0.2, 0.5, 1) for the respective curves, and b = 1, c = 6. 
The output from the sigmoid function is then used to rescale the baseline correlation parameter.
By construction
\[
\delta(x) = u(x)\,\delta_0
\]
which completes the $\delta$-based feedback loop between the controller and epidemic dynamics.

\clearpage
\bibliography{references}

\end{document}